\documentstyle[12pt,epsf]{article} 
\newcommand{\be}{\begin{equation}} 
\newcommand{\en}{\end{equation}}
\newcommand{\bea}{\begin{eqnarray}}
\newcommand{\ena}{\end{eqnarray}}

\newcommand{\hbo}{\hbox to 1 true cm {\hfill } }

\def\dslash{\partial\kern-.5em\slash}
\def\kslash{k\kern-.5em\slash}
\def\pslash{p\kern-.5em\slash}
\def\bs{\indent\indent}

\begin{document} 
\vglue 1truecm
  
\vbox{ UNITU-THEP-14/1996 
\hfill August 20, 1997 
}
  
\vfil
\centerline{\large\bf On the possibility of the dual Meissner effect } 
\centerline{\large\bf     induced by instantons } 
  
\bigskip
\centerline{ Kurt Langfeld } 
\vspace{1 true cm} 
\centerline{ Institut f\"ur Theoretische Physik, Universit\"at 
   T\"ubingen }
\centerline{D--72076 T\"ubingen, Germany.}
\bigskip
\vskip 1.5cm

\begin{abstract}
\noindent 

The classical Yang-Mills equation of motion is numerically 
investigated in the Lorentz gauge for a SU(2) gauge group. 
The color-electric field of two point-like charges is studied 
in the ''empty'' vacuum and in a state with an instanton present. 
The major effect for a fixed orientation of the instanton is 
that the color-electric field lines are expelled or attracted 
from the instanton region depending on the orientation of the 
instanton. If over the orientations of the instanton is averaged, 
this effect drops out. In this case of a random instanton orientation, 
we find that the external color-electric field is expelled from the 
instanton core. The origin of this effect is discussed.

\end{abstract}

\vfil
\hrule width 5truecm
\vskip .2truecm
\begin{quote} 
\end{quote}
\eject
\section{ Introduction } 
\label{sec:1}
\bs 

\noindent 
One of the most challenging problems in hadron physics nowadays is 
to understand the confinement of quarks in quantum-chromodynamics (QCD), 
which is the theory of strong interactions. From large scale numerical 
investigations of lattice Yang-Mills theory~\cite{wi74,eng87}, we 
know that QCD yields confinement, since Wilson's area law is satisfied. 
This implies that a linear rising confinement potential is formed 
between static color sources. 
Unfortunately, the numerical efforts have not yet revealed the 
basic mechanism of confinement. The knowledge of the qualitative 
mechanism would help to construct effective hadron theories the predictive 
power of which is not limited by unphysical quark thresholds~\cite{ja92}. 
Only effective quark models which heuristically incorporate the confinement 
are so far available. I would like to mention the Global Color 
Models~\cite{rob94} which realize confinement either by avoiding 
the pole at the real axis of the quark propagator or by infra-red 
slavery. In~\cite{la96a} it was proposed that quark confinement is due 
to random quark interactions induced by random gluonic background fields. 

A promising idea to understand the feature of quark confinement from 
first principles was given by 't~Hooft and Mandelstam. They propose 
that the so-called Abelian gauges are suited to understand the basic 
mechanism of confinement which is obscured in other gauges~\cite{tho76}. 
In these Abelian gauges magnetic monopoles emerge because of a residual 
U(1) gauge degree of freedom which is unfixed~\cite{kro87}. 
If these monopoles condense due to the Yang-Mills dynamics, a dual 
Meissner effect takes place implying that the color-electric flux is 
expelled from the vacuum. This scenario would naturally explain the linear 
rising potential between sources of color-electric flux. 
Subsequently, heuristic 
models were developed which explore the formation of an color-electric 
flux tube~\cite{ban79}. However, the dynamics which leads to the 
condensation of the monopoles is not understood so far. 

The original idea of 't~Hooft and Mandelstam has gained further support by 
a recent success~\cite{sei94} of Seiberg and Witten. 
They showed that the monopoles of $N=2$ (non-confining) super-symmetric 
Yang-Mills theory start to condense, if the theory is explicitly 
broken down to $N=1$ (confining) SUSY. One should, however, keep in mind 
that this scenario in super-symmetric Yang-Mills theory is 
quite different from that in standard QCD. Whereas in the previous case 
magnetic monopoles are present in the particle spectrum, magnetic 
monopoles are induced by gauge transformations in the latter case. 

The quasi-particles of standard Yang-Mills theory which show up as a 
solution of the classical Yang-Mills equation of motion are 
instantons~\cite{inst}. An implicit scheme to construct all 
instanton solutions was provided in~\cite{at77}. Novel explicit 
instanton configurations for a SU(N), $N\ge 3$ can be found 
in~\cite{re93}. Instantons are generally believed to play 
an important role in the QCD ground state. They may possibly trigger the 
spontaneous breaking of chiral symmetry~\cite{dya97} 
and offer an explanation of the $U_{A}(1)$ problem~\cite{tho76b}. 

Numerical investigations, based on a classical instanton interaction, 
show that instantons are strongly correlated 
in the Yang-Mills ground state implying that they occur as a 
liquid~\cite{sh82} rather than in a dilute gas phase~\cite{ca78}. 
Including quantum fluctuations, it was discovered that instantons 
possess a medium range attractive interaction which is solely due to 
the instability of the perturbative vacuum~\cite{la94}. 
The strong-coupling expansion within the field strength approach 
to Yang-Mills theory~\cite{sch90} suggests that the interaction 
due to quantum effects is strong enough to induce a condensation 
of instantons~\cite{la94b}. Whether the instantons exist in a crystal 
type structure or in a strongly correlated liquid is still beyond 
the scope of the present approaches. 

The question whether instantons are important for the dual Meissner 
effect has gained recent attention~\cite{che95,bro96,tep96,thu96,feu97}. 
In order to address this question, the profile of an instanton in 
Abelian gauges were investigated in continuum Yang-Mills 
theory~\cite{che95,bro96} as well as in the lattice 
version~\cite{tep96,thu96}. In~\cite{che95}, a monopole world line was 
observed which penetrates the center of the instanton. 
In contrast, the lattice simulations~\cite{tep96,thu96} have reported 
closed loops of monopoles 
which encircle the center of the instanton. The length of the monopole 
loop was investigated throughout the deconfinement transition. 
The discrepancy between the lattice and the continuum approach was 
clarified by Brower at al.~\cite{bro96}. They showed that the configurations 
which minimize the gauge fixing functional indeed lead to closed monopole 
world lines. The conclusion at hand is that large monopole loops 
play an important role for confinement. Recently, lattice calculations 
reported an enhanced probability for monopoles inside the 
instanton~\cite{feu97} indicating that instantons play a role in the 
context of the dual Meissner effect. 

In this paper, we will directly focus on this role of the instantons. 
For this purpose, we will numerically 
solve the SU(2) Yang-Mills equation of motion with two static color 
sources present. The color-electric flux stemming from the color sources 
will be investigated in the ''empty'' vacuum and in a vacuum where 
an instanton is present. Whether the external color-electric field 
lines are expelled or attracted depends on the orientation of the 
instanton relative to the charges. If over the orientation of the 
instanton is averaged, the picture changes qualitatively. We will 
provide evidence that in the latter case the external field is 
repulsed from the instanton core. 

The paper is organized as follows: in the following section, 
we briefly review some aspects concerning the relation 
of instantons and confinement. In section \ref{sec:2}, 
the Meissner effect in solid state physics is addressed 
for a later comparison with the situation in SU(2) Yang-Mills theory. 
The numerical approach to solve the classical Yang-Mills equation of motion 
is introduced in section \ref{sec:3}. It is shown 
that this approach nicely reproduces an instanton configuration. 
In subsection \ref{sec:3.3}, two point-like color-electric charges 
are added. The electric field lines stemming from these charges are 
investigated in an ''empty'' vacuum (subsection \ref{sec:3.3}) 
and in a state with an instanton present (subsection \ref{sec:3.4}). 
The case of a random instanton orientation is presented in 
section \ref{sec:3b}. 
Conclusions as well as an argument to understand qualitatively the numerical 
results of the latter section are left to the final section.

\section{ Instantons imply confinement? }
\label{sec:1a} 

\noindent 
In this section, we briefly review some results in the literature 
which address the question how relevant instantons are for 
the confinement mechanism. 

This question has a long 
history~\cite{aha78,shu78,pol88}. A semi-classical evaluation of the 
Yang-Mills partition function by Aharonov et al.~\cite{aha78} 
has indicated that instantons support the existence of infinite long 
flux tubes (fluxon). These strings with non-vanishing flux appear as 
fluctuations on top of the instanton background. It was shown that an 
ensemble of these random-walking ''fluxons'' gives rise to Wilson's area 
law. In this scenario, the impact of the instantons on confinement 
is secondary. 

Considering fluctuations around an instanton, 't~Hooft showed 
that large size instantons receive a large weight in the 
Yang-Mills partition function~\cite{tho76b}. Expectation values of 
condensates are even ill-defined due to the contribution of these large 
size instantons. Subsequently, it was argued by Shuryak that the presence 
of quarks and gluons cure the large size instanton problem~\cite{shu78}. 
He showed that the presence of a non-vanishing quark and/or gluon density 
provides a natural cutoff on the instanton radius. As a consequence, 
the presence of quarks and gluons yield an increase in the vacuum energy 
density implying that they should be expelled from an instanton 
dominated vacuum. This picture naturally supports the MIT bag model 
of hadronic matter. Note that we will focus at the classical level when we 
will elaborate below the response of the instantons on the external charges. 

In their pioneering work, Polikarpov and Veselov provided evidence 
that the instantons play a minor role for confinement~\cite{pol88}. 
They studied the emergence of instantons 
in SU(2) lattice Yang-Mills theory, when a equilibrium lattice configuration 
is cooled down. They extracted the ratio $\chi $ of the 
string tension of the configuration {\it after} cooling with the one of the 
lattice equilibrium configuration as function of the topological charge 
present on the lattice. One crucial observation is that $\chi $ 
increases with increasing topological charge $Q$. For instance, 
$\chi $ reaches $20 $\% for a sufficiently large Wilson loop and for 
$Q=3$. However, large topological charges rarely appear on the lattice. 
The value of $\chi $ averaged over the instanton vacua which appear 
by freezing out equilibrium configurations is of order $5 $\%. 

The outcome of the latter considerations is twofold: an instanton medium 
which emerges from the freezing of a lattice equilibrium configuration 
cannot give rise to the full string tension. Fluctuations on top of 
the instanton medium seem to play an important role for confinement. 
Recent lattice simulations indicate that these fluctuations might be 
magnetic strings, which are attached to the instanton~\cite{feu97}.  
Secondly, lattice equilibrium configurations are dominated by 
fluctuations rather than by the classical solutions of the Yang-Mills 
equation of motion. The later fact is confirmed by recent studies 
in~\cite{thu96}. The authors investigated the 
evolution of the topological charge $Q$ during the cooling process. 
They found that the lattice equilibrium configurations are dominated 
by configurations with $Q=0$. However, if these configurations are 
cooled down, configurations with larger $Q$-values emerge. This 
demonstrates that the fluctuations on the lattice (at realistic 
values of the coupling strength) are large enough to change the 
topological charge $Q$. 

In this paper, we will elaborate a detailed picture of the 
response of an instanton to the presence of static charges 
on a ''microscopic'' level. We will thereby provide new information 
to clarify the role of the instantons for the dual Meissner effect rather 
than to estimate the importance of the (semi-) classical vacuum for the 
true ground state of Yang-Mills theory.

\section{ The Meissner effect of solid state physics } 
\label{sec:2} 
\bs

\noindent 
In 1933 Meissner and Ochsenfeld discovered that super-conductors expel 
magnetic flux. The microscopic ingredient which leads to this effect is 
the condensation of electric charge, which is carried by pairs of 
electrons~\cite{ash81}. For later comparison with the Yang-Mills case, we 
briefly review this effect by considering a Landau-Ginzburg 
theory~\cite{lan50} which is described in terms of the Lagrangian 
\be 
{\cal L} \; = \; 
\vert (\partial _\mu \, + i {\cal A}_\mu (x) ) \, \phi(x) \vert ^2 
\; + \; \frac{1}{4e^2} {\cal F}_{\mu \nu }^2 \; - \; j_\mu (x) 
{\cal A}_\mu (x) \; + \; V(\phi ^2) \; . 
\label{eq:1} \; , 
\en 
where ${\cal F}_{\mu \nu }$ is the electro-magnetic field strength 
built from the gauge potential ${\cal A}_\mu $, and $e$ is the electric 
charge. $V(\phi^2)$ is the tree-level Higgs-potential, which is minimal 
for a non-vanishing value $\phi ^2 = \phi _0^2$ providing a condensation 
of the scalar field. $j_\mu (x) $ is the external current. 
The Lagrangian is invariant under U(1) gauge transformations. 
The scalar field $\phi (x)$ transforms homogeneously under these 
transformations and therefore carries electric charge. 

\begin{figure}[t]
\parbox{6cm}{ 
\hspace{1cm} 
\centerline{ 
\epsfxsize=6cm
\epsffile{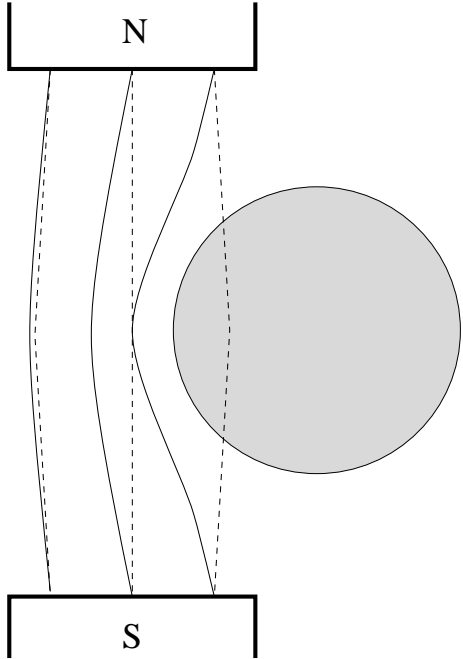} 
}
} \hspace{1cm}
\parbox{7cm}{ 
\caption{ The Meissner effect: magnetic field lines with a 
   super-conducting cylinder present (solid lines) and without the 
   cylinder (dashed lines). } 
} 
\label{fig:1} 
\end{figure} 
In order to mimic the scenario of a super-conductor, we assume that 
the scalar field forms a homogeneous condensate $\phi_0$ of electric 
charge, which breaks the U(1) gauge symmetry. The classical 
equation of motion can be easily obtained from (\ref{eq:1}), i.e. 
\be 
\partial _{\mu } {\cal F}_{\mu \nu } (x) \; - \; 
2 e^2 \phi _0^2 \, {\cal A}_\nu (x) \; = \; - j_\nu (x) \; . 
\label{eq:2} 
\en 
In the gauge $\partial _\mu {\cal A}_\mu (x) =0$, this equation 
can be transformed into an Helmholtz equation for the magnetic field 
${\cal B}_i$. In the absence of external currents, this equation 
becomes 
\be 
\partial ^2 {\cal B}_i(x) \; - \; 2e^2 \phi _0^2 \, {\cal B}_i(x) 
\; = \; 0 \; . 
\label{eq:3} 
\en 
This is the key equation to understand the Meissner effect. Eq.(\ref{eq:3}) 
tells us that the magnetic field exponentially decreases inside 
a super-conductor with slope given by the strength of the condensate 
$\phi _0$. Figure 1 qualitatively shows the behavior 
of the magnetic field lines, if a super-conducting cylinder is present. 
The main observation is that the magnetic field lines are expelled 
from the super-conducting region. 

Let us compare London's equation (\ref{eq:2}) with 
the classical equation of motion of an SU(2) Yang-Mills theory. 
This equation can be derived from the Euclidean action which is 
defined in the appendix, i.e. 
\be 
\partial _\mu F^a_{\mu \nu } [A] (x) \; - \; \epsilon ^{abc} 
F^c_{\nu \mu }[A] \, A^{b}_{\mu } (x) \; = \; - g^2 \, j^a_{\nu }(x) \; , 
\label{eq:4} 
\en 
where $\epsilon ^{abc}$ is the anti-symmetric tensor in 3 dimensions, and 
$j^a_{\nu }(x)$ are the external currents. The 
field strength, i.e. 
\be 
F^a_{\mu \nu }[A] \; = \; \partial _\mu A^{a}_\nu \, - \, 
\partial _\nu A^{a}_\mu \, + \, \epsilon ^{abc} A^b_\mu A^c_\nu \; , 
\label{eq:5} 
\en 
is constructed from the non-Abelian gauge field $A^a_\mu (x)$. 
The external source $j^a_\mu $ is a real function of space-time and can 
be understood 
as the current generated by Euclidean quark fields (see (\ref{eq:a5})). 
We will adopt the Lorentz gauge, $\partial _\mu A^a_\mu (x) =0 $, and 
will study the impact of this current on instantons throughout this 
paper. 

For a vanishing external current $(j^a_\nu=0)$, eq.(\ref{eq:4}) 
allows for non-trivial solutions, which are 
known as instantons~\cite{inst}. In the gauge $\partial _\mu A_\mu^a (x) 
=0 $, their gauge field and the corresponding field strength is 
given by 
\be 
A^{a \, inst}_{\mu } (x) \; = \; \eta ^a_{\mu \nu } x_\nu \, 
\frac{ 2 }{ x^2 + \rho ^2} \; , \hbo 
F^{a \, inst }_{\mu \nu }(x) \; = \; - \, \eta ^a_{\mu \nu } \, 
\frac{ 4 \rho ^2 }{ (x^2 + \rho ^2)^2 } \; , 
\label{eq:6} 
\en 
where $\eta ^a_{\mu \nu }$ are the anti-symmetric 't~Hooft symbols, i.e. 
\be 
\eta ^a_{0i} \; = \; \delta _{ai} \; , \hbo 
\eta ^a_{ik} \; = \; \epsilon ^{aik} \; , \; \; \; 
i,k = 1 \ldots 3 \; , 
\label{eq:7} 
\en 
and $\rho $ is the radius of the instanton. Instantons therefore 
correspond to spots of non-vanishing field strength. 

Let us now briefly discuss the linear response of the gauge field, i.e. 
$a^a_\mu (x) $, to the external source $j^a_\nu (x)$. 
Decomposing $A_\mu ^a (x) = A^{a \, inst}_{\mu } (x) + a^a_\mu (x) $, 
we assume that the current is sufficiently weak implying that one 
can expand the e.o.m.~(\ref{eq:4}) to linear order in $a^a_\mu (x)$. 
If we confine us to the region close to the center of the 
instanton, i.e. $\sqrt{x^2} \ll \rho $, the e.o.m.\ becomes 
\be 
\partial ^2 a^a_\nu (x) \, - \, \frac{3}{2} \epsilon ^{abc} F^c_{\nu \mu } 
\left[A^{inst}(x=0)\right] \, a^b_\mu (x) \; = \; - g^2 \, j^a_\nu (x) ; . 
\label{eq:7a} 
\en 
Since the matrix $\epsilon ^{abc} F^{c}_{\mu \nu }$ possesses negative 
eigenvalues, eq.~(\ref{eq:7a}) describes screening of the field strength 
as well as anti-screening depending on the color index $a$ under 
consideration. 
The field strength $F^a_{\mu \nu }$ at the instanton center sets the 
scale of the (anti-) screening length and therefore plays a similar role as 
the scalar field $\phi _0^2$ in (\ref{eq:2}). We learn from 
eq.~(\ref{eq:7a}) that Meissner type effects occur due to the 
non-linear nature of the Yang-Mills e.o.m., if instantons are present in 
the ground state of SU(2) Yang-Mills theory. The crucial difference 
is, however, that these effects lead to a repulsion or an attraction 
of the color-electric field lines depending on the orientation 
of the instanton. 

In this paper, we will further investigate these effects resorting 
to numerical methods in order to go beyond the linear response 
theory. We will provide evidence that the electric flux produced by the 
charges is expelled from the instanton core, if we average over the 
instanton orientation. In this case, the above leading order effect obtained 
by the linear response approach drops out.

\newpage 
\section{ The dual Meissner effect of SU(2) Yang-Mills theory } 
\label{sec:3} 

\subsection{ The numerical approach } 
\label{sec:3.1} 
\bs 

\noindent 
In the following, we will work in the Lorentz gauge 
\be 
\partial _\mu A^{a}_{\mu }(x) \; = \; 0 \; . 
\label{eq:8} 
\en 
We will numerically investigate the solutions of the 
classical equation of motion (\ref{eq:4}) (we set $g^2=1$ for simplicity)
of SU(2) Yang-Mills theory, 
which is a non-linear second order partial differential equation. 
To this aim, we discretize the 4-dimensional Euclidean space-time 
on a grid consisting of $31^4$ points, and replace derivatives by the 
corresponding differences, e.g. 
\be 
\partial _\mu f(x) \rightarrow \frac{1}{2h} ( f(x+h \hat{e}_\mu ) 
\, - \, f(x-h \hat{e}_\mu ) ) \; , 
\label{eq:9} 
\en 
where $\hat{e}_\mu $ is the unit vector in $\mu $-direction, and where 
$h$ is the grid spacing. The discretized partial differential 
equation is then solved by iteration. For definiteness, we have chosen 
Neumann boundary conditions. 
We have checked that the impact of the boundaries on the 
field configurations under consideration is small (see below). 

The approach should not be confused with the lattice version of 
Yang-Mills theory~\cite{wi74}. The 
grid is only a mathematical tool to solve the partial differential 
equation. For our purposes, we need not worry about local 
gauge invariance, but confine ourself to the definite gauge 
choice (\ref{eq:8}). 

\begin{figure}[t]
\parbox{6cm}{ 
\hspace{1cm} 
\centerline{ 
\epsfxsize=8cm
\epsffile{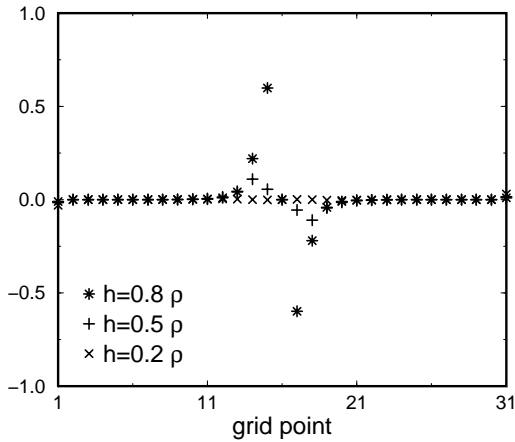} 
}
} \hspace{1cm}
\parbox{7cm}{ 
\caption{ The ''$0$'' produced from the discretized version of the 
   e.o.m.~(\protect{\ref{eq:4}}) by inserting the instanton 
   configuration (\protect{\ref{eq:6}}). } 
}
\label{fig:2} 
\end{figure} 
In order to get a first idea of the error due to the discretization 
(and to check the set up of the program), we cast the analytically 
known instanton configuration $A^{a \, inst}_{\mu }(x)$ (\ref{eq:6}) 
onto the grid, and checked to what extent the discretized version 
of the l.h.s.~of (\ref{eq:4}) reproduces the zero at the r.h.s.~(note 
that the external source $j^a_\mu $ is set to zero at the moment). Choosing 
$x_{2,3,4}=0$ (this selects a line which passes the center of the instanton) 
and e.g.~$a=2$ and $\nu =3$, the deviation from zero is shown in figure 
2 as function of $x_1$ for several values of the grid 
spacing $h$. The result should be compared with the intrinsic 
scale which is set by the maximum field strength of the instanton 
($4$ in units of the instanton radius). The most important observation 
is that the result becomes significantly better, if $h$ is decreased. 
This confirms that the discretization works correctly.

\subsection{ The instanton on the grid } 
\label{sec:3.2} 
\bs 
\vspace{-.5cm} 

\begin{figure}[t]
\centerline{ 
\epsfxsize=8cm
\epsffile{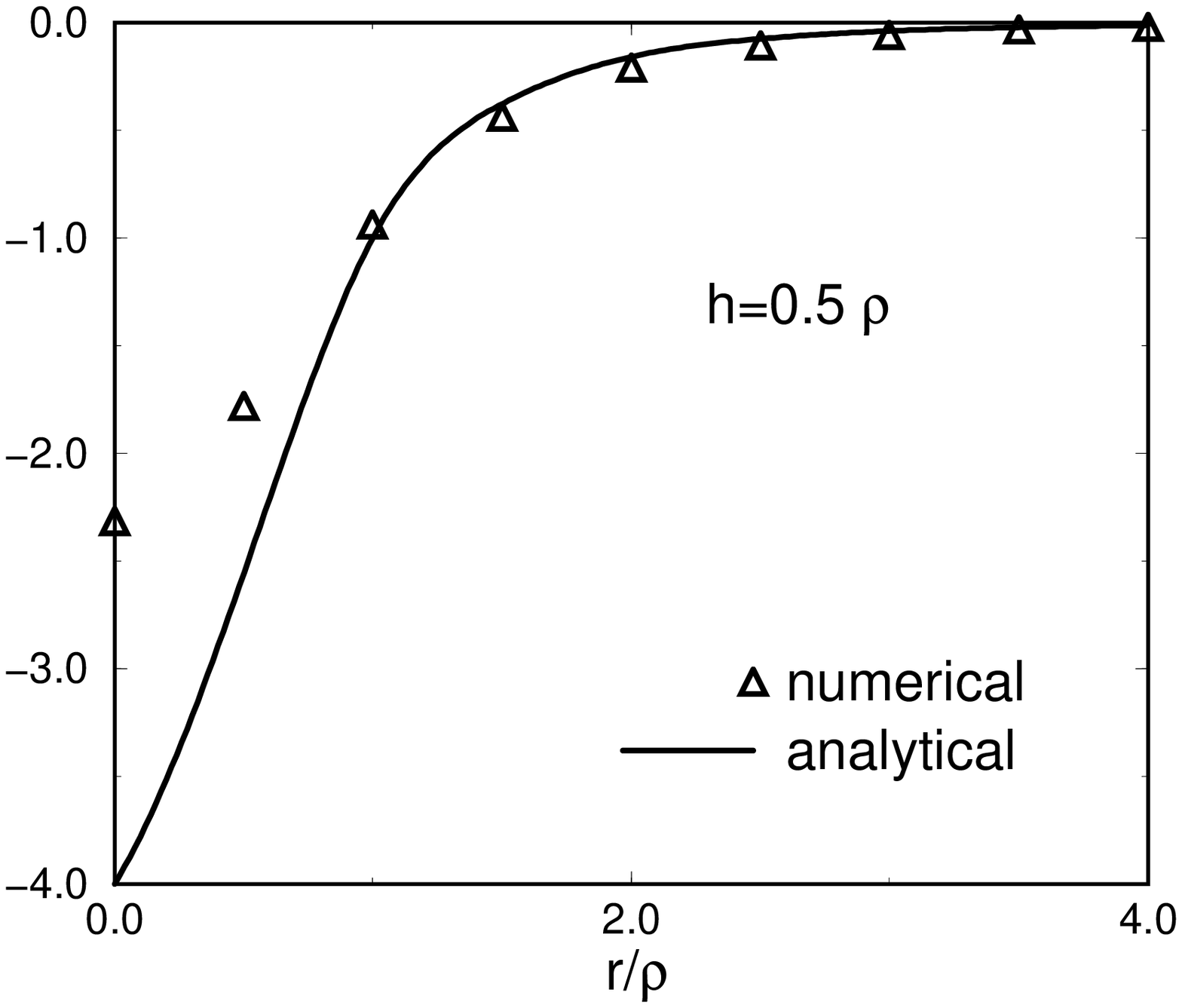} 
\epsfxsize=8cm
\epsffile{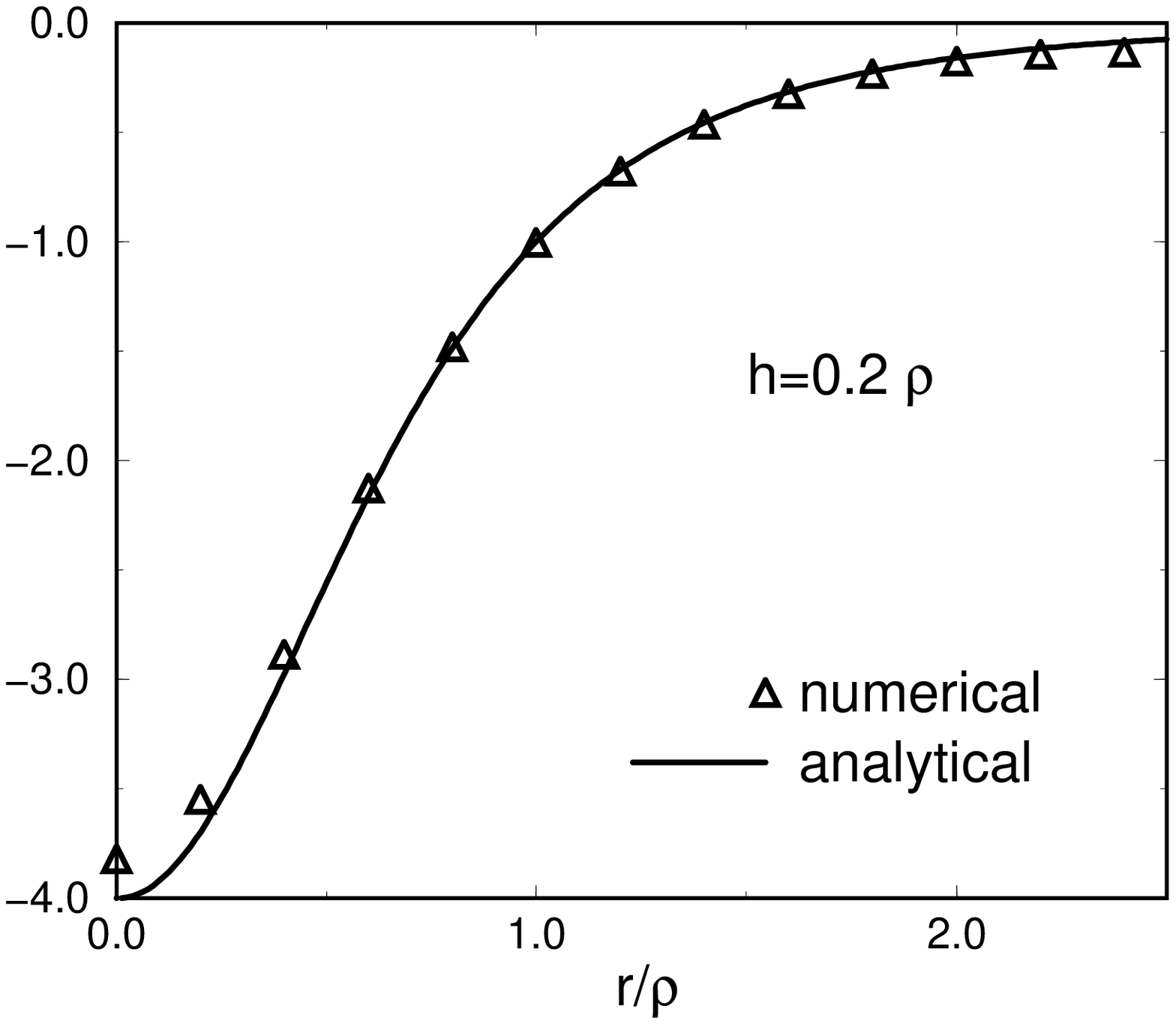} 
}
\vspace{-.8cm} 
\caption{ The color-electric field $E^1_1(r=x_1)$ of an instanton 
   for $x_{2,3,4}=0$: 
   The numerical solution (triangles) in comparison with the 
   analytic result (\protect{\ref{eq:6}}). } 
\label{fig:3} 
\end{figure} 
\noindent 
In this subsection, we study the grid version of an instanton 
configuration, which emerges as a solution of the discretized 
e.o.m.\ (\ref{eq:4}) at zero external source $(j^a_\mu (x) =0)$. 
The SU(2) instanton possesses eight zero-modes, 
since the instanton configuration breaks symmetries of the Lagrangian. 
Four translational, three rotational and one dilatational degrees of 
freedom do not lead to a change of the action. In the 
pseudo-particle description of the instanton, these zero-modes 
give rise to collective coordinates, which 
must be fixed in order to arrive at a definite configuration. 
For this purpose, it is convenient to employ a 
procedure which minimizes the action by the method of steepest 
decent. Then zero-modes do not contribute to the gradient configuration 
which is added in each iteration step to the actual profile in order to 
further reduce the action. This implies that the starting configuration 
of the iteration process completely fixes the collective coordinates of 
the instanton. Here we have chosen the analytic profile (\ref{eq:6}) as 
the starting configuration. 

\newpage 
\centerline{ 
\epsfxsize=11cm
\epsffile{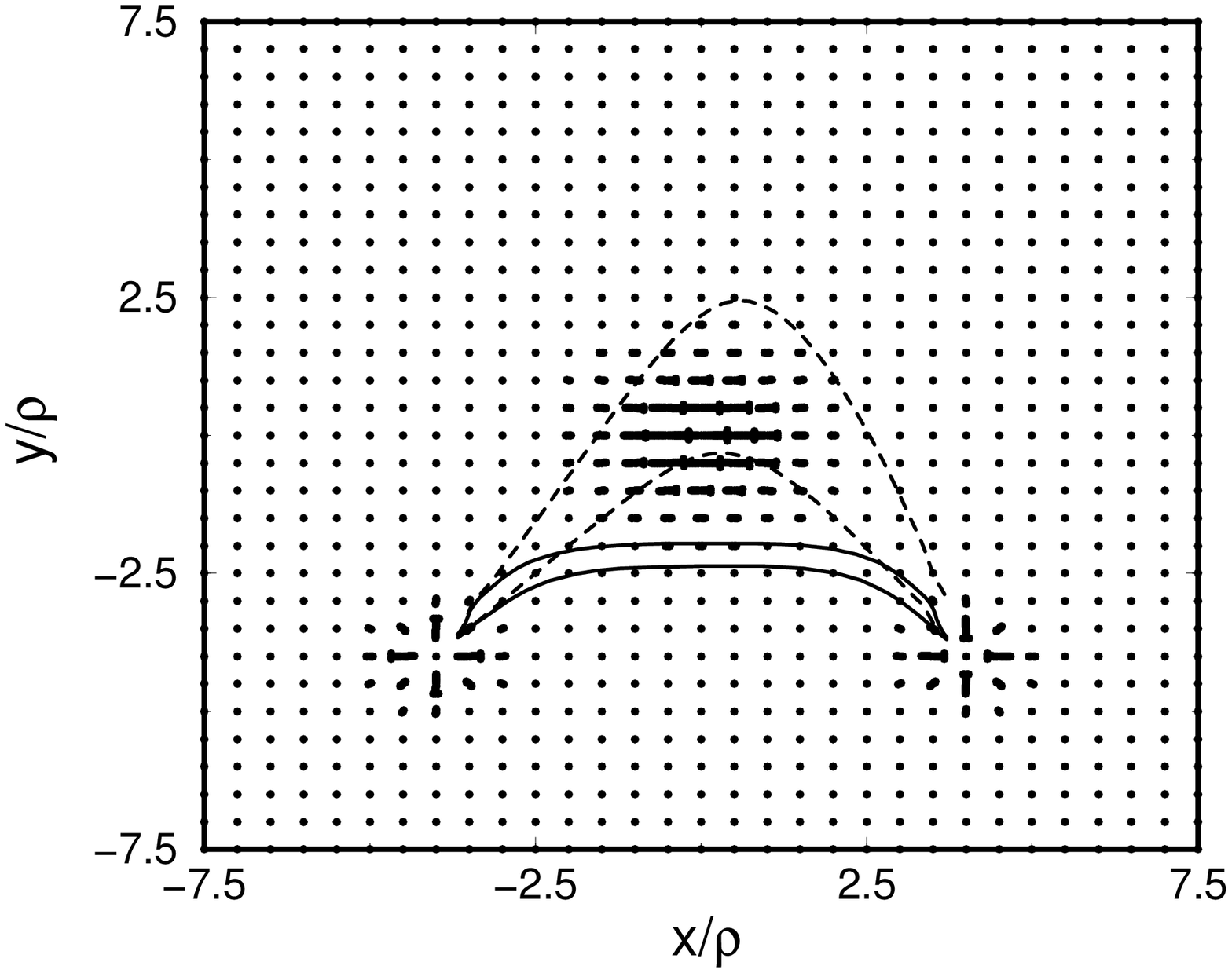} 
}
\centerline{ 
\epsfxsize=11cm
\epsffile{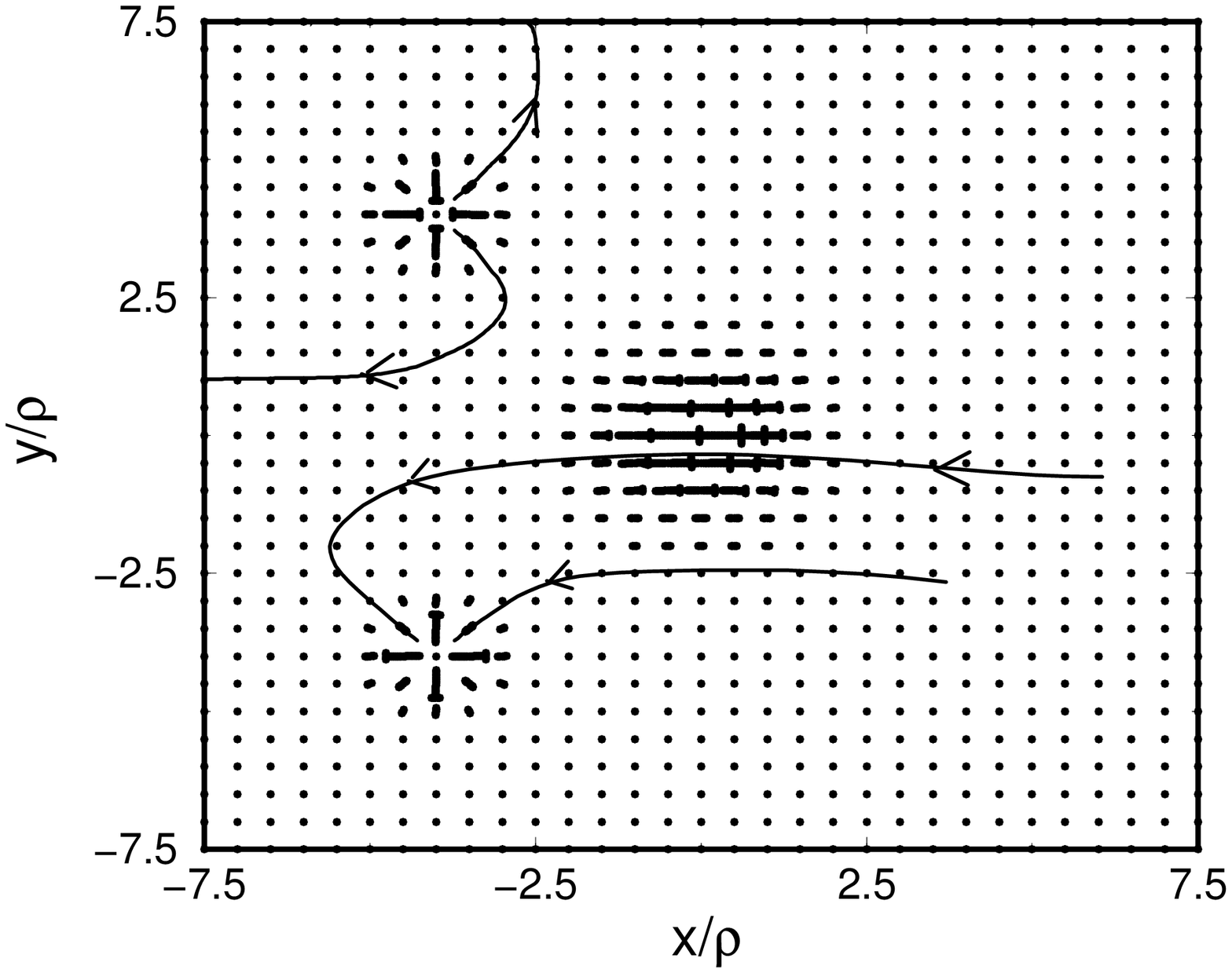} 
}
\begin{figure}[b]
\caption{ The color-electric field $\vec{E}^1$ in the $xy$-plane 
   produced by two color-electric charges in a state with one instanton 
   present. The dashed lines indicate the field lines of the charges 
   in an empty vacuum. }
\label{fig:4} 
\end{figure} 
\newpage 

After convergence was obtained, we have calculated the 
color-electric field of our grid configuration with the help of a 
discretized version of the field strength tensor (\ref{eq:5}). 
Choosing the color direction 
$a=1$, we learn from (\ref{eq:6}) and (\ref{eq:7}) that the 
only non-vanishing color-electric field points in $\hat{e}_1$ 
direction. We then have compared the numerical result for the space-time 
dependence of this color-electric field, i.e. $E^1_1(x)$, with the analytic 
form (\ref{eq:6}). The result in shown in figure \ref{fig:3}. 
For $h=0.5 \, \rho $, the field strength in the center of the 
instanton is somewhat underestimated, whereas the result is satisfactory 
for $h=0.2 \, \rho $. However, the influence of the boundaries 
becomes visible at the edges of the grid in the latter case.

\subsection{ Electric charges in empty space } 
\label{sec:3.3} 
\bs 

\noindent 
In this subsection, we study the color-electric field produced by static 
color sources in non-Abelian SU(2) gauge theory. 
For this purpose, we have chosen the external current at the 
the r.h.s.\ of the equation of motion (\ref{eq:4}) to be 
\be 
j^a_\mu (x) \; = \; \delta _{a1} \, \delta _{\mu 0} \; 
\left[ \delta ^{(3)} (\vec{x}-\vec{x}_1) \, - \, 
\delta ^{(3)} (\vec{x}-\vec{x}_2) \right] 
\; . 
\label{eq:10} 
\en 
This choice corresponds to two static color-electric charges of 
color direction $a=1$ which are located at the spatial positions 
$\vec{x}_1$ and $\vec{x}_2$. 

In order to get a feeling of the color-electric flux produced by these 
sources, we calculated the color-electric field $\vec{E}^{a=1}$ using 
an empty space as starting configuration of the iteration process, 
which solves the partial differential equation (\ref{eq:4}). 
To be specific, we have chosen 
\be 
\vec{x}_1 \; = \; (4\rho , -4\rho , 0) \; , \hbo 
\vec{x}_2 \; = \; (-4\rho , -4\rho , 0) \; , 
\label{eq:11} 
\en 
where $\rho $ is an arbitrary length scale at the moment and will be 
the instanton radius later. If only sources for one color direction 
is present, the non-linear terms of the Yang-Mills equation of motion 
(\ref{eq:4}) drop out, and the e.o.m.\ essentially reduces to a 
Maxwell equation in the relevant color channel. For a later comparison 
with the case with an instanton present, we numerically calculated 
the color-electric 
field $\vec{E}^1$ in the $xy$-plane at each grid point. The 
color-electric field lines show the behavior anticipated from the 
analogy to classical electrodynamics.

\subsection{ Electric charges in the instanton background } 
\label{sec:3.4} 
\bs

\noindent 
The picture changes drastically, if the vacuum contains an instanton 
configuration. We here only consider sufficiently weak external charges 
which do not completely deform the instanton and which therefore 
preserve the pseudo-particle character of the instanton. 
In this limit, we are interested in the interplay of the color-electric 
field strengths of the external charges and the instanton as function 
of the collective coordinates. Note that in the presence of the external 
charges, the collective coordinates of an instanton in empty space 
do not correspond to zero-modes anymore, but a particular choice of these 
parameters exists, if the action is minimal. 
One possibility to constrain the collective coordinates is to include 
the term 
\be 
\lambda \int d^4 x \; \left( F^a_{\mu \nu } [A_{ex}] - 
F^a_{\mu \nu }[A] \right) ^2 
\label{eq:11a} 
\en 
to the action. The field $A_{ex}$ is thereby the gauge potential 
of a single instanton in empty space with a definite choice of 
origin, orientation and radius. The parameter $\lambda $ acts like 
a Lagrange multiplier. For sufficiently small values of $\lambda $, 
the motion of the instanton along the (former) zero-mode directions 
due to the external source is blocked, and the collective coordinates 
are dictated by the background field. Using several (small) values of 
$\lambda $, one verifies that the field strengths configuration, produced 
by the instanton and the charges, remains (almost) unchanged by the 
additional term (\ref{eq:11a}). 

In order to be specific, the iteration process which minimizes the action 
was started with the analytic instanton configuration (\ref{eq:6}) 
(which is also used as the background field in (\ref{eq:11a})) 
casted onto the grid. 
The upper picture of figure \ref{fig:4} shows the result for $t=0$, 
where the instanton develops its maximum strength. 
The color-electric field lines of the instanton, which point from the 
right to the left, are clearly visible. The length of the vectors 
on the grid points also give a rough impression of the 
instanton field strength profile (\ref{eq:6}). Some color-electric 
field lines are also shown (solid lines). For comparison, the 
color-electric field lines of the case without the instanton are 
shown as dashed lines, too. A field line is specified by the 
angle between the field line and the line connecting the two charges. 
This makes it possible to compare the situation with and without 
instanton. Our main observation is that the color-electric field lines 
of the charges are clearly expelled from the region where the instanton 
is located. 

\begin{figure}[t]
\parbox{6cm}{ 
\hspace{1cm} 
\centerline{ 
\epsfxsize=8cm
\epsffile{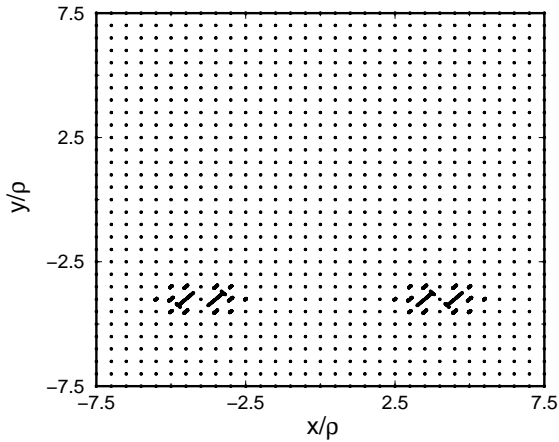} 
}
} \hspace{1cm}
\parbox{7cm}{ 
\caption{ The color-electric field $\vec{E}^1$ in the $xy$-plane of the 
   charge instanton system with the color-electric field of the unperturbed 
   instanton subtracted. } 
} 
\label{fig:5} 
\end{figure} 
Let us study this effect, if the location of the two 
charges is changed to 
\be 
\vec{x}_1 \; = \; (-4\rho , -4\rho , 0) \; , \hbo 
\vec{x}_2 \; = \; (-4\rho , 4\rho , 0) \; . 
\label{eq:12} 
\en 
The result is shown in the lower picture of figure \ref{fig:4}. 
The upper charge is the source of the color-electric field, the lower 
charge is the drain. One observes that the color-electric field lines 
which spread out the upper charge are pushed back from the instanton, 
whereas the lines of the drain charge are pulled in. The net effect 
of the instanton is to align the external electric field lines with 
its internal orientation. 

One might argue that the deformation of the color-electric field lines 
of the charges in the presence of the instanton is solely due 
to the superposition of the instanton field and the field from the 
charges. Figure 5 shows that this is not the case. In this 
picture, we have subtracted the color-electric field of a pure instanton 
(without any charges present) 
from the full color-electric field. If the total field is simply 
a superposition, one should recover the Maxwell type field lines
(this is what would happen in classical electrodynamics). 
The result of this subtraction is completely different from the 
field distribution expected from classical electrodynamics. 
This is possible due to the non-linear nature of the classical equation of 
motion (\ref{eq:4}).

\section{ Random instanton orientation } 
\label{sec:3b} 
\bs 

\noindent 

In the previous section, we have investigated the distribution 
of the electric field strength, produced by a charge 
anti-charge pair, for a definite choice of the instanton orientation. 
No evidence for the dual Meissner effect was found so far, since 
the external color-electric field lines are attracted or pulled back 
from the instanton core depending on the orientation. 
Here, we will study whether a net repulsion of the color-electric field 
lines takes place, if the we average over the instanton orientation. 
For this purpose, we will 
calculate the space-time distribution the field strength squared 
$F^a_{\mu \nu } F^a_{\mu \nu }$, where we average over the orientation 
of the instanton and keep its position fixed. 
The case of the random instanton orientation is interesting from a 
physical point of view for the following reason: a realistic description 
of the Yang-Mills ground state resorts to a liquid of 
instantons~\cite{sh82}. In this liquid, the orientation of a particular 
instanton is arbitrary. The small amounts of action which are necessary 
to move the instanton along its former (i.e. single instanton) zero-mode 
directions are provided by the entropy of the medium. 

\begin{figure}[t]
\parbox{6cm}{ 
\hspace{1cm} 
\centerline{ 
\epsfxsize=8cm
\epsffile{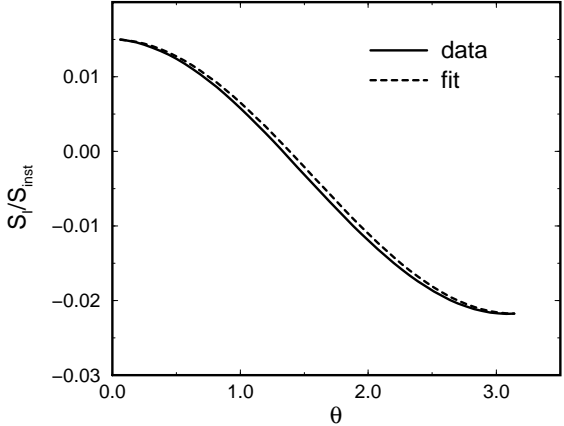} 
}
} \hspace{1cm}
\parbox{7cm}{ 
\caption{ The instanton interaction $S_I$ as function of 
    the instanton orientation $\theta $. } 
}
\label{fig:6} 
\end{figure} 
Let us first study the dependence of the instanton interaction with the 
external charges on the orientation $O^{ab}$ of the instanton, i.e. 
\be 
A^{a \, inst}_{\mu } (x) \; = \; O^{ab} \eta ^b_{\mu \nu } x_\nu \, 
\frac{ 2 }{ x^2 + \rho ^2} \; , \hbo 
O \; = \; \left( \begin{array}{ccc} 
  \cos \theta  &  - \sin \theta & 0 \cr 
  \sin \theta  &    \cos \theta & 0 \cr 
   0           &         0      & 1 \cr 
                 \end{array} \right) \; , 
\label{eq:b1} 
\en 
where $O^{ab}$ is a particular orthogonal matrix parameterized by the angle 
$\theta $. Since we will switch on electric charges in color $a=1$ 
direction, it is sufficient for our purposes to study matrices $O^{ab}$ 
which rotate the instanton color-electric field in the $x_1 x_2$-plane only. 
The instanton interaction $S_I$ is defined by 
\be 
S_I \; = \; S_{IC} \, - \, S_{inst} \, - \, S_{charges} \; , 
\label{eq:b2} 
\en 
where $S_{IC}$ is the action of the instanton-charge system, and 
$S_{inst}$ and $S_{charges}$ is the action of a single instanton and 
the charges, respectively. If the charges are weak and far located 
from the instanton origin, the interaction $S_I$ can be calculated 
in an elegant way~\cite{ca78}, i.e. 
\be 
S_I^0 \; \propto \; \eta ^a_{\mu \nu } F^{a \, ext.}_{\mu \nu } \; , 
\label{eq:b3} 
\en 
where $F^{a \, ext.}_{\mu \nu }$ is the field strength of the 
external charges at the instanton origin, and where the superscript 
indicates that it was assumed that the this field strength is 
slowly varying over the instanton region. If we align 
the charges in $\hat{e}_1$-direction symmetric to the $x_2$-axis, 
their color-electric field at the instanton center is also oriented 
in $\hat{e}_1$ direction. This implies that 
$S_I^0 \propto \cos \theta $ for the particular choice of orientation 
(\ref{eq:b1}). The numerical result for $S_I$ is shown in figure 6. 
The strengths of the charges actually yield 
$S_{charges} / S_{inst} = 6.7 \% $. The distance of one charge to the 
instanton center was $1.7 \, \rho $. A fit according $S_I \approx 
a \, \cos \theta \, + \, b$ is also shown in figure \ref{fig:6}. 
A $\cos \theta $-modulation of the data, as predicted by (\ref{eq:b3}), 
is clearly visible. In addition, one observes a non-vanishing offset $b$ 
towards negative values. This offset can be understood as follows: 
if we put the field configurations of the charges and of the 
unperturbed instanton on the grid as starting configuration, 
the method of steepest decent further reduces the action in order to 
finally arrive at a solution of the Yang-Mills equations of motion. 
Since action of the starting configuration does not depend of the 
instanton orientation, the angle average of $S_I$ is negative. 
From this argument, it is obvious that the offset $b$ cannot obtained in 
the linear response approach, which is the basis of eq.~(\ref{eq:b3}). 

\begin{figure}[t]
\centerline{ 
\epsfxsize=12cm
\epsffile{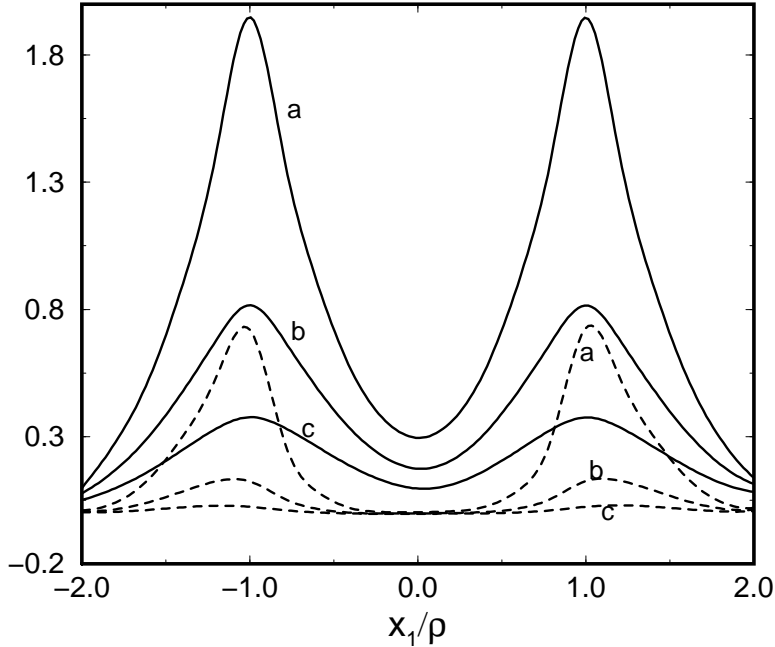} 
}
\vspace{-.5cm} 
\caption{ $(F^{a}_{\mu \nu } F^{a}_{\mu \nu })^{1/2} 
    (x_1,x_2,x_{0,3}=0)$ for several values of $x_2$ (cases a-c) 
    of the charges in empty space (solid) and of the instanton-charge 
    system with $(F^2)^{1/2}$ of an unperturbed instanton subtracted 
    (dashed). In the case of the instanton-charge system, we have 
    averaged over the instanton orientations. } 
\label{fig:7} 
\end{figure} 
Finally, we will average over the $\theta $-angle, which defines the 
instanton orientation with respect to the charges. We will study the 
variation of the field strength squared, i.e. $F^a_{\mu \nu } 
F^a_{\mu \nu }(x)$, in space-time rather than the integrated quantity $S_I$. 
Since $F^2$ of a single instanton decreases like $1/r^8$ for large distances 
$r$ from the instanton center and since $F^2$ of a single charge 
asymptotically behaves like $1/r^4$, it is more convenient for illustration 
purposes to study the quantity $\sqrt{F^2}$. In the present investigation, 
we have located the charges at 
\be 
\vec{x}_1 \; = \; (\rho , -\rho , 0) \; , \hbo 
\vec{x}_2 \; = \; (-\rho , -\rho , 0) \; , 
\label{eq:b4} 
\en 
Figure 7 shows $\sqrt{F^2}$ of the 
charge anti-charge pair (solid lines) as function of $x_1$ $(x_{3,0}=0)$ 
for several values of $x_2$, i.e. case $a$ $x_2=-0.4 \, \rho $, 
case $b$ $x_2=-0.2 \, \rho $ and case $c$ $x_2=0$. 
This result is compared with $\sqrt{F^2}_{cI}$ produced by the 
instanton charge configuration (dashed lines), where $\sqrt{F^2}$ of a 
single instanton was subtracted and where over the orientations of the 
instanton was averaged. The latter quantity is therefore a measure of the 
field strength which exists on top of that of an unperturbed instanton. 
One clearly observes that 
$\sqrt{F^2}$ produced by the external charges is suppressed at the 
space time region occupied by the instanton. In order to quantify this result, 
we introduce the ratio 
\be 
\kappa (x_2) \; = \; \frac{ \left[ \sqrt{F^2}^{\, peak}_{charges} (x_2) 
\, - \, \sqrt{F^2}^{\, peak}_{cI}(x_2) \right] 
}{ \sqrt{F^2}^{\, peak}_{charges}(x_2) } \; , 
\en 
where $\sqrt{F^2}^{\, peak}_{charges} $ and $\sqrt{F^2}^{\, peak}_{cI}$ are 
the maximum values of $\sqrt{F^2}$ of the charges in empty space and of 
the subtracted charge instanton system, respectively, at a given line 
$(x_2, x_{3,0}=0)$. The ratio $\kappa $ directly measures the 
suppression of the external color-electric field inside the instanton. 
The numerical calculation reveals the following result
$$
\hbox{ \hfill 
\begin{tabular}{|l|l|l|l|l|} \hline 
$x_2 \; \; \; $ & $0$ & $-0.2$ & $-0.4$ & $-0.6$ \\ \hline 
$\kappa $       & $92$\% & $84$\% & $63$\% & $28$\% \\ \hline 
\end{tabular} 
\hfill } $$ 
One finds that a net suppression of the external field inside the 
instanton occurs, if over the instanton orientation is averaged. The 
suppression increases towards the instanton center.

\section{ Discussions and conclusions } 
\label{sec:4} 
\bs

\noindent 
Among many other things, Polikarpov and Veselov studied the 
instanton medium which emerges in lattice Yang-Mills theory, 
if an equilibrium configuration is cooled down~\cite{pol88}. They 
found in particular, that this cooled medium exhibits a non-vanishing 
string tension, which, however, is only 5\% of the full string tension. 
Whether this residual confinement is due to some excitations still present 
on top of the instantons at the cooled lattice or due to an intrinsic 
property of the instanton medium is not clear yet. 

In order to provide new information concerning the interesting question 
whether the instantons play a role for the dual Meissner effect, we have 
studied the impact of an instanton on the 
color-electric flux produced by two static, point-like color-electric 
charges. The classical equation of motion of a SU(2) gauge theory was 
numerically solved in Lorentz gauge. The color-electric 
field lines of the charges were investigated in detail in an ''empty'' 
vacuum as well as in a state where an instanton is present. We found that 
the external color electric field lines are expelled or attracted 
depending on the orientation of the instanton relative to the charges. 
The net effect is that the instanton aligns the external field lines 
which is in accordance with the result of the leading order of the 
linear response theory~\cite{ca78}. However, if we average over the 
orientations of the instanton, this leading order effect drops out. 
In sub-leading order, a substantial repulsion of the color-electric field 
lines from the instanton core region survives. 

The numerical outcome can be qualitatively understood as follows: 
in an ''empty'' space, the electric flux spreading out the charges 
distributes over the whole space in order to avoid large 
concentrations of field strength which amounts for a large action. 
If only charges of unique color are present, the scenario is the same for 
the Abelian (classical electro-dynamics) and the non-Abelian case. 
However, the non-linearity of the Yang-Mills equation of motion (in 
fact the topological structure of the theory) allows for non-trivial 
configurations, i.e. instantons, which are local minima of the Yang-Mills 
action. These configuration must exhibit a definite strength of 
the color-electric and color-magnetic field (see (\ref{eq:6})) in order 
to minimize the action. Any distortion of the instanton field strength, 
e.g. produced by the presence of external charges, yields a sudden 
increase of the action. This is the reason that the color-electric 
field is expelled from the instanton core region, once the dominant dipole 
interaction of the instanton with the charges drops out due to a 
random instanton orientation. 

Whether this effect is a pre-cursor of quark confinement giving rise 
to a non-vanishing string tension, if the Yang-Mills ground state 
is modeled as a strongly correlated medium of randomly oriented 
instantons, is an interesting question, which is left to future studies.

\bigskip 
\noindent 
{\bf Acknowledgments: } 

\noindent 
I thank E.~V.~Shuryak for useful informations and interesting 
comments, and Mannque Rho for helpful remarks. I am also indebted 
to H.~Reinhardt for many discussions on the issue of monopole 
condensation in Yang-Mills theories as well as support.

\vspace{1 true cm} 
\noindent 
{\bf \large Appendix:  Euclidean gauge theory }

\bigskip 
The Euclidean partition function of a SU(2) gauge theory, fermions 
included, is described by the functional integral 
\be 
Z \; = \; \int {\cal D} q \; {\cal D} q^\dagger \; {\cal D} A^a_\mu \; 
\exp \left\{ - \int d^4x \; \left[ 
\frac{1}{4g^2} F^a_{\mu \nu } F^a_{\mu \nu } + q^\dagger 
(i\dslash + im - A^a_\mu t^a ) q \right] \right\} \; , 
\label{eq:a1} 
\en 
where the gauge fields $A_\mu ^a$ are considered as real fields, and 
where the fermion fields $q$ transform under the fundamental representation 
of the $SU(2)$ gauge group which is spanned by the generators $t^a$. 
$g$ is the gauge coupling constant and $m$ is current quark mass. 
The action is invariant, if the fields transform under a global 
O(4) symmetry, i.e. 
\be 
A^{a\, \prime }_\mu \; = \; \Lambda _{\mu \nu } A^a_\nu \; , 
\hbo 
q' \; = \; S(\Lambda ) q \; , 
\label{eq:a2} 
\en 
which is the pendant to the Lorentz symmetry in Minkowski space. 
The orthogonal matrices $\Lambda _{\mu \nu }$ and the matrices 
$S(\Lambda )$ are spanned by the anti-symmetric generators 
$\omega _{\mu \nu }$, e.g. $ \Lambda _{\mu \nu } \; = \; [ \exp 
\{ - \omega \} ] _{\mu \nu } $. The matrices $S(\Lambda )$ 
satisfy 
\be 
S(\Lambda ) \gamma _\mu S^\dagger (\Lambda ) \; = \; \Lambda _{\mu \nu } 
\gamma _\nu \; , 
\label{eq:a3} 
\en 
where $\gamma _\mu $ are the hermitian Dirac matrices which fulfill 
$\{\gamma _\mu , \gamma _\nu \} = 2 \delta _{\mu \nu }$. 
From (\ref{eq:a3}), it is obvious that 
\be 
j^a_\mu (x) \; = \; q^{\dagger }(x) \, \gamma _\mu \, t^a \, q(x) 
\label{eq:a5} 
\en 
is a hermitian current which transforms like a O(4) vector. 
The zeroth component of this vector is the charge density which is the 
subject of investigations in this paper. 

Finally, let us check that the coupling of the current (\ref{eq:5}) 
to the gauge fields $A^a_\mu $ is correctly chosen, i.e. a gauge 
transformation does not induce imaginary parts in $A^a_\mu $. 
One easily observes that the transformation 
\be 
q^\prime (x) \; = \; U(x) q(x) \; , \hbo 
U(x) \in SU(2) 
\en 
leaves the action invariant, if the gauge fields transform according 
\be 
t^a A^{a \, \prime }_\mu (x) \; = \; U(x) \, t^a A^a_\mu (x) \, 
U^\dagger (x) \; - \; i U(x) \partial _\mu U^\dagger (x) \; . 
\label{eq:a6} 
\en 
It is easy to verify that the transformation (\ref{eq:a6}) leaves the 
gauge field $A^a_\mu $ real. The contribution of the interaction 
of the quark current and the gauge fields to the Euclidean action 
is therefore 
\be 
S_{int} \; = \; - \int d^4x \; j^a_\mu (x) A^a_\mu (x) \; . 
\en 
It is straightforward to derive the equation of motion (\ref{eq:4}) 
from the action in (\ref{eq:a1}) by taking the functional derivative 
with respect to the gauge fields $A^a_\nu (x)$. 

We here provide the right hand side of the equation of motion 
(\ref{eq:4}), i.e. the current, and calculate the gauge fields. 
First information on the gauge fields can be obtained by taking 
the divergence of the equation of motion. Using the e.o.m., a direct 
calculation yields 
\be 
\partial _\nu j^a _\nu (x) \; = \; \epsilon ^{abc} A^b_\mu (x) 
j^c_\mu (x) \; . 
\label{eq:a7} 
\en 
In this paper, we study the response of the gauge fields to static 
charges, i.e. 
\be 
j_\mu ^a (x) \; = \; \delta _{\mu 0} 
\sum _i \rho_i ^a \, \delta (\vec{x} - \vec{x_i}) \; . 
\en 
Eq.~(\ref{eq:a7}) then tells us that the zeroth component of the 
gauge fields at the position of the charges either vanishes, i.e. 
$A_0 ^a (t,\vec{x_i}) =0$, or that 
the gauge fields are oriented parallel to the charge color vector, i.e. 
$A^a_0 (t,\vec{x_i}) \propto \rho ^a_i $.

\end{document}